\documentclass{PoS}
\rightline{\parbox{4cm}{\large\rm
Edinburgh 2006/32\\
SHEP-0632
}}

\usepackage{amssymb}
\usepackage{amsmath}

\newcommand{\be}{\begin{equation}}
\newcommand{\ee}{\end{equation}}
\def\ltorder{\mathrel{\raise.3ex\hbox{$<$}\mkern-14mu
 \lower0.6ex\hbox{$\sim$}}}

\title{$K\to\pi l\nu$ form factor with $N_f=2+1$ dynamical domain wall
  fermions}

\ShortTitle{$K\to\pi l\nu$ form factor}

\author{D.~J.~Antonio,$^a$ P.~A.~Boyle,$^a$ C.~Dawson,$^b$
  T.~Izubuchi,$^{b,c}$ A.~J\"uttner,$^d$
  C.~Sachrajda,$^d$ S.~Sasaki,$^{b,e}$ A.~Soni,$^f$ R.~J.~Tweedie,$^a$
  J.~M.~Zanotti$^a$\thanks{Speaker (jzanotti@ph.ed.ac.uk).}\\
  \llap{$^a$}School of Physics, University of Edinburgh,
  Edinburgh EH9 3JZ, UK\\
  \llap{$^b$}RIKEN-BNL Research Center, Brookhaven National
  Laboratory,
  Upton, NY 11973, USA\\
  \llap{$^c$}Institute for Theoretical Physics, Kanazawa University,
  Kanazawa, Ishikawa 920-1192, Japan\\
  \llap{$^d$}School of Physics and Astronomy, University of
  Southampton,
  Southampton, SO17 1BJ, UK\\
  \llap{$^e$}Department of Physics, University of Tokyo, Tokyo
  113-0033,
  Japan\\
  \llap{$^f$}Physics Department, Brookhaven National Laboratory,
  Upton,
  NY 11973, USA\\
  \vspace*{2mm} {\rm \bf UKQCD/RBC Collaboration} }

\abstract{We present the latest results from the UKQCD/RBC
  collaborations for the $K_{l3}$ form factor with $2+1$ flavours of
  dynamical domain wall quarks. Simulations are performed on
  $16^3\times 32\times 16$ and $24^3\times 64\times 16$ lattices with
  three values of the light quark mass, allowing for an extrapolation
  to the chiral limit.
  After interpolating to zero momentum transfer, we obtain the
  preliminary result $f_+^{K\pi}(0)=0.9680(16)$, which is in excellent
  agreement with an earlier $N_f=2$ result.}

\FullConference{XXIVth International Symposium on Lattice Field Theory\\
                July 23-28, 2006\\
                Tucson, Arizona, USA}

\begin{document}

\section{Introduction}
\vspace*{-2mm}

$K\rightarrow \pi l\nu\ (K_{l3})$ decays provide an excellent avenue
for an accurate determination of the Cabibbo-Kobayashi-Maskawa (CKM)
\cite{Cabibbo:1963yz} quark mixing matrix element, $|V_{us}|$.
This is done by observing that the decay amplitude is proportional to
$|V_{us}|^2 |f_+(q^2)|^2$, where $f_+(q^2)$ is the form factor
defined from the $K\to \pi$ matrix element of the weak vector current,
$V_\mu=\bar{s}\gamma_\mu u$
\be
\langle \pi(p^\prime) \big | V_\mu \big | K(p)\rangle = (p_\mu +
p_\mu^\prime) f_+(q^2) + (p_\mu - p_\mu^\prime) 
f_-(q^2),\ \ q^2=(p-p^\prime)^2\ .
\label{eq:ME}
\ee

In chiral perturbation theory (ChPT), $f_+(0)$ is expanded in terms of
the light pseudoscalar meson masses, $m_\pi,\,m_K,\,m_\eta$
\be
f_+(0) = 1 + f_2 + f_4 + \ldots,\quad (f_n={\cal
  O}(m^n_{\pi,\,K,\,\eta}))\ .
\ee
Conservation of the vector current ensures that $f_+(0) = 1$ in the
$SU(3)$ flavour limit, while $SU(3)$ flavour breaking effects
occur only at second order in $(m_s-m_{ud})$ due to the Ademollo-Gatto
Theorem \cite{Ademollo:1964sr}, which states that $f_2$ receives no
contribution from local operators appearing in the effective theory.
As a result, $f_2$ can be determined unambiguously in terms of
$m_\pi$, $m_K$ and $f_\pi$, and takes the value $f_2=-0.023$ at the
physical masses \cite{Leutwyler:1984je}.

Our task is now reduced to one of finding
\be
\Delta f = f_+(0) - (1 + f_2) \ .
\label{eq:deltaf}
\ee
In order to obtain a result for $f_+(0)$ which is accurate to
$\sim$1\%, it is sufficient to have a 20-30\% error on $\Delta f$.
Until recently, the standard estimate of $\Delta f=-0.016(8)$ was due
to Leutwyler \& Roos \cite{Leutwyler:1984je}, however a more recent
ChPT analysis favours a positive value, $\Delta f=0.007(12)$
\cite{Cirigliano:2005xn}.
A calculation of $\Delta f$ on the lattice is therefore essential.

The last few years have seen an improvement in the accuracy of lattice
calculations of this quantity
\cite{Becirevic:2004ya,Okamoto:2004df,Tsutsui:2005cj,Dawson:2005zv},
with the results favouring a negative value for $\Delta f$ in
agreement with Leutwyler \& Roos.
The most recent study used 2 flavours of dynamical domain wall
fermions to obtain a result $\Delta f = -0.009(9)$
\cite{Dawson:2006qc}.

The UKQCD and RBC collaborations have embarked on a program to improve
on earlier studies by using $N_f=2+1$ flavours of dynamical domain
wall fermions at light quark masses and on large volumes.
We present here preliminary results from this study.

\section{Lattice Techniques}
\vspace*{-2mm}
\subsection{Parameters}
\vspace*{-2mm}

We simulate with $N_f=2+1$ dynamical flavours generated with the
Iwasaki gauge action \cite{Iwasaki:1985we} at $\beta=2.13$, which
corresponds to an inverse lattice spacing $a^{-1}\approx
1.6\,\text{GeV}$ ~\cite{Rob}, and the domain wall fermion action
\cite{Kaplan:1992bt} with domain wall height $M_5=1.8$ and fifth
dimension length $L_s=16$.
This results in a residual mass of $am_\text{res}\approx 0.00308(3)$
\cite{Rob}.
The simulated strange quark mass, $am_s=0.04$, is very close to it's
physical value \cite{Rob}, and we choose three values for the light
quark masses, $am_{ud}=0.03,\,0.02,\,0.01$, which correspond to pion
masses $m_\pi\approx 630$, 520, 390 MeV \cite{Rob}.
The calculations are performed on two volumes, $16^3\times 32$ and
$24^3\times 64$, at each quark mass.
For more simulation details, see \cite{Rob}.

\subsection{$f_o(q^2_\text{max})$}
\label{sec:f0q2max}

We start by rewriting the vector form factors given in
Eq.~(\ref{eq:ME}) to define the scalar form factor
\be
f_0(q^2) = f_+(q^2) + \frac{q^2}{m_K^2 - m_\pi^2}f_-(q^2)\ ,
\ee
which can be obtained at $q^2_\text{max}=(m_K^2 - m_\pi^2)$ with high
precision from the following ratio \cite{Hashimoto:1999yp}
\be
R^1(t',t) = \frac
{C^{K\pi}_4(t',t;\vec{0},\vec{0})C^{\pi K}_4(t',t;\vec{0},\vec{0})}
{C^{KK}_4(t',t;\vec{0},\vec{0})C^{\pi\pi}_4(t',t;\vec{0},\vec{0})}
\xrightarrow[t,(t'-t) \to \infty]{} \frac{(m_K +
  m_\pi)^2}{4m_K m_\pi} \big| f_0(q^2_\text{max})\big|^2\ ,
\label{eq:ratio1}
\ee
where the three-point function is defined as
\be
C_{\mu}^{PQ}(t', t,\vec{p}\,',\vec{p})  
   = \sum_{\vec{x}, \vec{y}}
   e^{-i\vec{p}\,'(\vec{y}-\vec{x})} e^{-i\vec{p} \vec{x}} 
\big < 0 \big |  {\cal O}_Q\big |
   Q \big > \big < Q \big |
   V_\mu \big | P \big > \big < P \big | 
   {\cal O}^\dagger_P \big | 0 \big > \ ,
\ee
with $P,Q=\pi$ or $K$ and ${\cal O}_{\pi(K)}$ is an interpolating
operator for a pion(kaon).
We note that $R^1(t',t)=1$ in the $SU(3)_\text{flavour}$ symmetric
limit, hence any deviations from unity are purely due to
$SU(3)_\text{flavour}$ symmetry breaking effects.

\begin{figure}[tb]
\vspace*{-7mm}
\includegraphics[width=6cm,angle=-90]{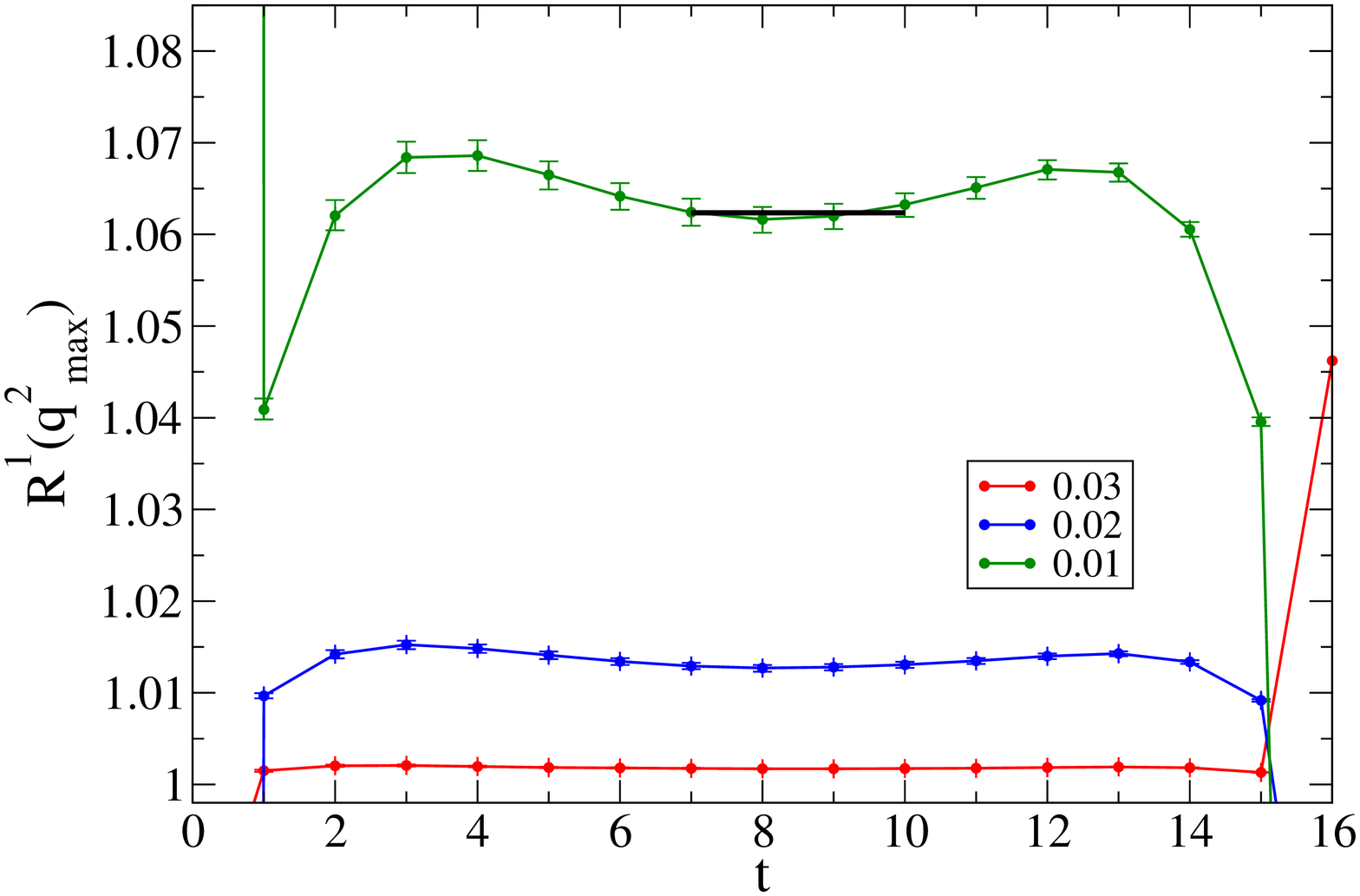}
\includegraphics[width=6cm,angle=-90]{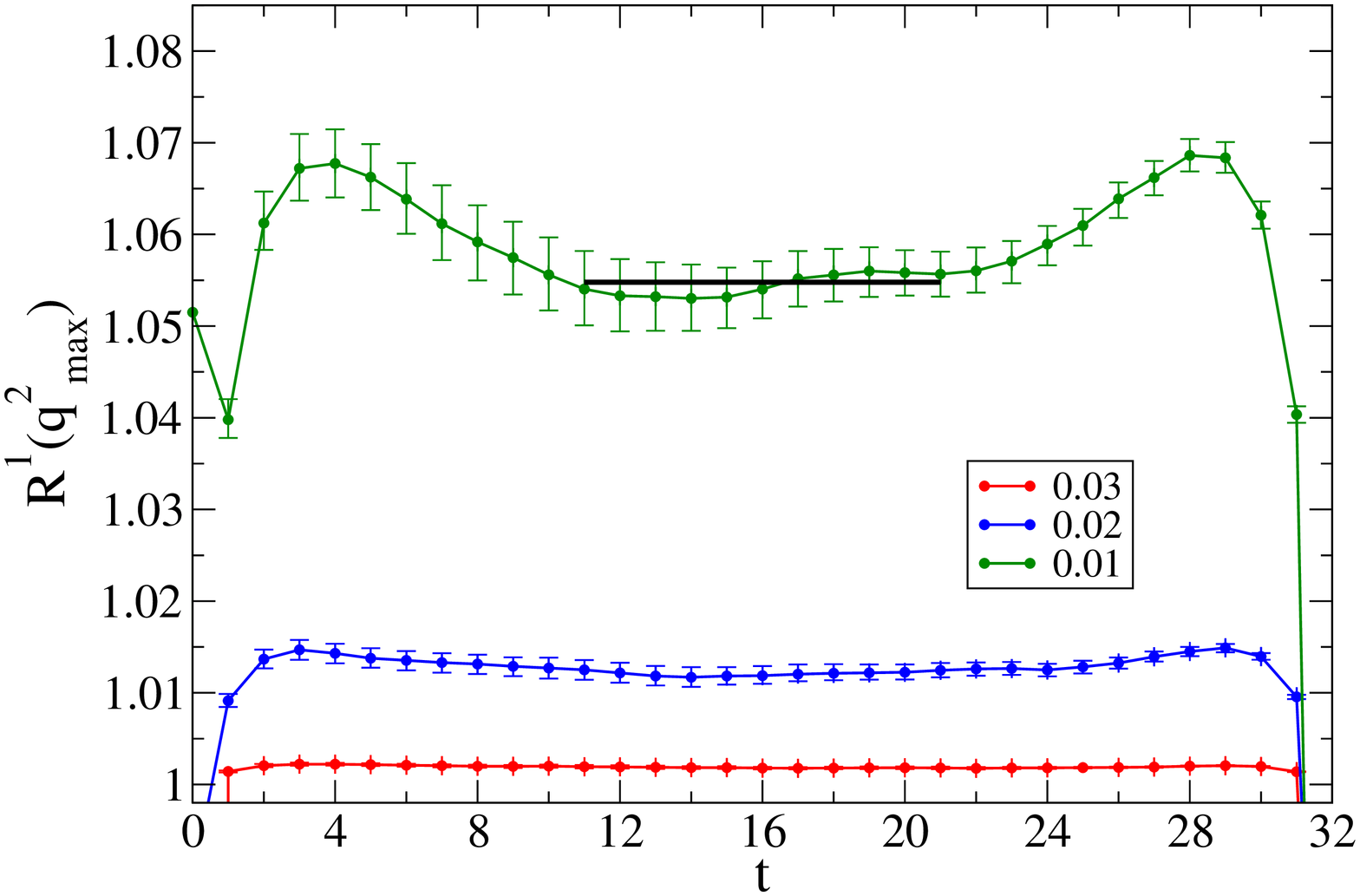}
\vspace*{-6mm}
\caption{Ratio for $f_0(q_\text{max}^2)$, $R^1(t',t)$, as defined in
  Eq.~({\protect\ref{eq:ratio1}}), for three simulated light masses
  $am_{ud}=0.03,\,0.02,\,0.01$ for two different volumes, $16^3\times 32$
(left) and $24^3\times 64$ (right). Further simulation parameters can be
found in ~{\protect\cite{Rob}}.}
\label{fig:ratio1}
\vspace*{-3mm}
\end{figure}

In the left (right) plot of Fig.~\ref{fig:ratio1} we display our
results for $R^1(t',t)$ for each of the simulated quark masses as
obtained on the $16^3\times 32$ ($24^3\times 64$) lattices.
It is immediately obvious that $R^1(t',t)$ can be measured with a very
high level of statistical accuracy.
We also note that the ratio becomes larger the further we move away
from the $SU(3)_\text{flavour}$ limit.
Since there is no spatial momentum involved in this ratio, the results
obtained on the two different volumes should agree, and any difference
can only be due to finite size effects.
The two plots in Fig.~\ref{fig:ratio1} indicate that within
statistical errors, finite size effects on $f_0(q^2_\text{max})$ are
negligible. 

Finally, we note that the increased time extent of the $24^3\times 64$
lattice allows for a longer plateau from which we can extract our
result.

\subsection{Investigating the momentum transfer dependence}

To study the $q^2$ dependence of $f_0(q^2)$, we construct the second
ratio
\be
R^2(t',t;\vec{p}\,',\vec{p}) = \frac
{C^{K\pi}_4(t',t;\vec{p}\,',\vec{p})C^K(t;\vec{0})C^\pi(t'-t;\vec{0})}
{C^{K\pi}_4(t',t;\vec{0},\vec{0})C^K(t;\vec{p})C^\pi(t'-t;\vec{p}\,')}
\xrightarrow[t,(t'-t) \to \infty]{} \frac{E_K(\vec{p}) +
  E_\pi(\vec{p}\,')}{m_K + m_\pi}
F(p',p)\ ,
\label{eq:ratio2}
\ee
from which we are able to extract $f_+(q^2)$ via
\be
F(p',p) = \frac{f_+(q^2)}{f_0(q^2_\text{max})} 
\left( 1 + \frac
{E_K(\vec{p}) - E_\pi(\vec{p}\,')}
{E_K(\vec{p}) + E_\pi(\vec{p}\,')} 
\xi(q^2)\right)\,,\ 
\xi(q^2) = \frac{f_-(q^2)}{f_+(q^2)}\ .
\label{eq:Fpp}
\ee
$C^{\pi(K)}(t;\vec{p})$ in Eq.~(\ref{eq:ratio2}) is the standard pion
(kaon) two point function.

\begin{figure}[tb]
\vspace*{-7mm}
\includegraphics[width=6cm,angle=-90]{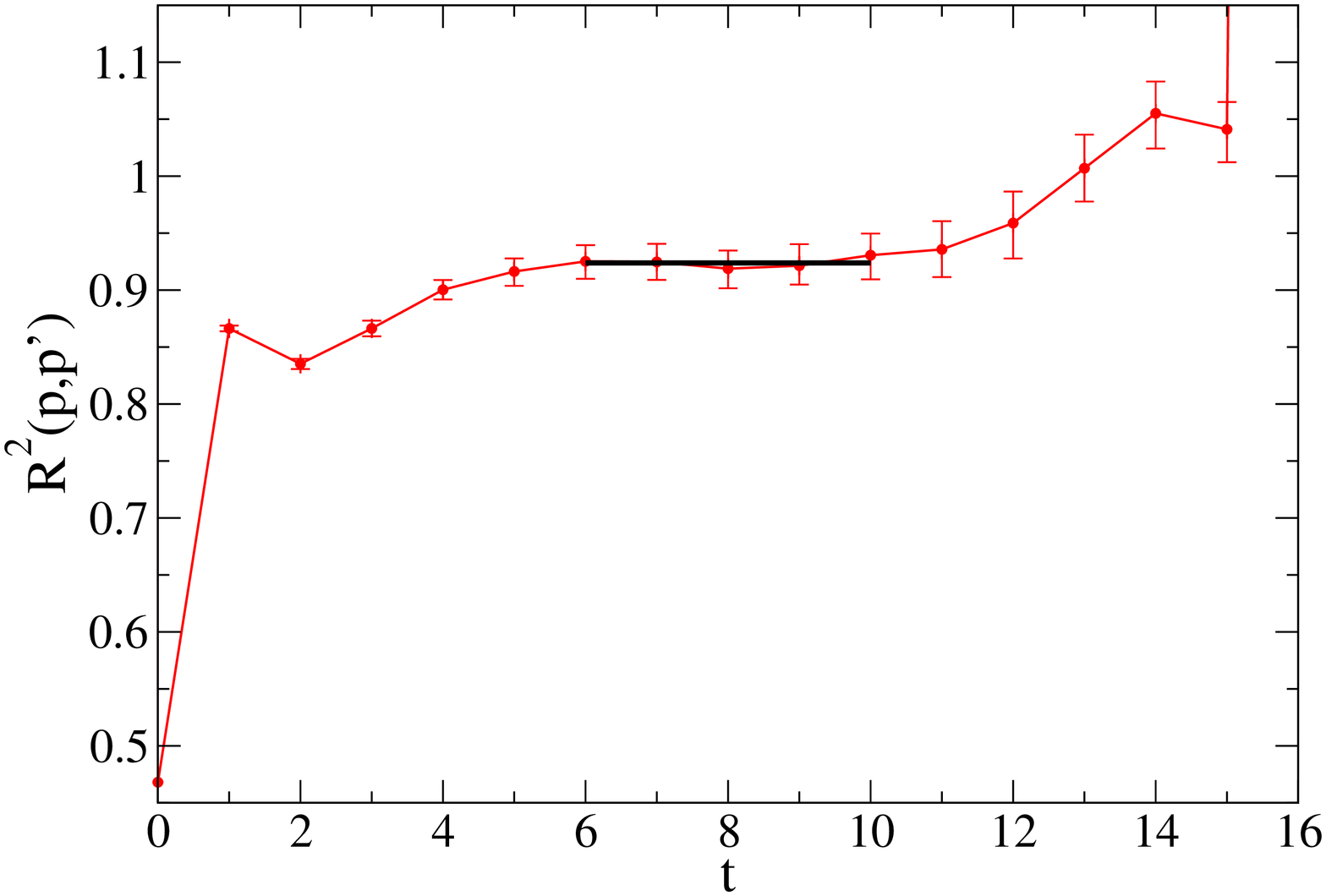}
\includegraphics[width=6cm,angle=-90]{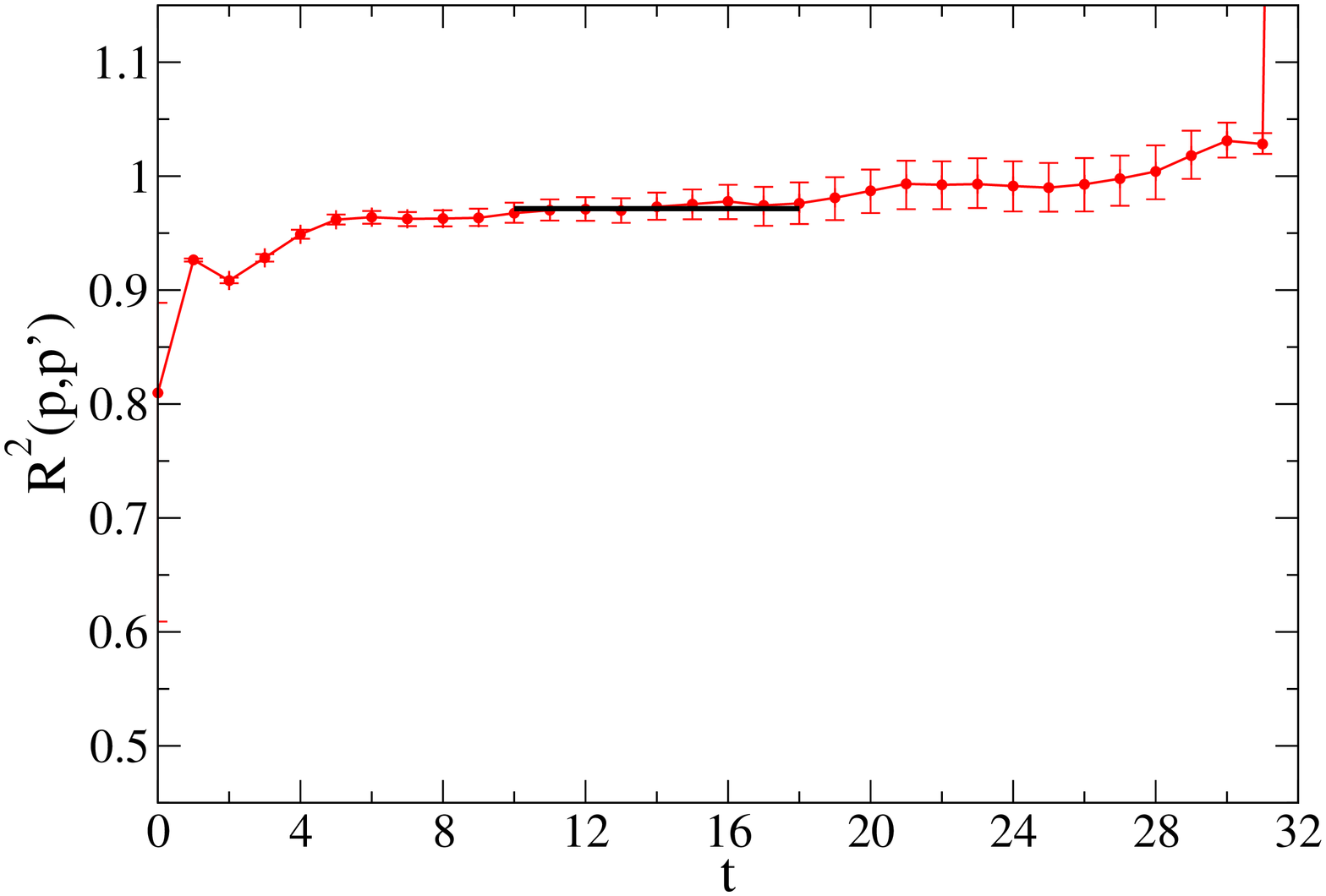}
\vspace*{-6mm}
\caption{Ratio for $F(p,p')$, $R^2(t',t;\vec{p}\,',\vec{p})$, as defined in
  Eq.~({\protect\ref{eq:ratio2}}), for bare quark mass $am_{ud}=0.02$
  with momentum transfer, $|a\vec{q}|^2=1$ for two different volumes,
  $16^3\times 32$ (left) and $24^3\times 64$ (right).}
\label{fig:ratio2}
\vspace*{-3mm}
\end{figure}

Figure~\ref{fig:ratio2} displays a typical example of
$R^2(t',t;\vec{p}\,',\vec{p})$ for bare light quark mass
$am_{ud}=0.02$ and momentum transfer $|a\vec{q}|^2=1$, where we have
set $\vec{p}\,'=0$ and averaged over equivalent 3-momenta
corresponding to the same 4-momentum transfer.
The left plot shows our result from the $16^3\times 32$ lattice, while
the right displays $24^3\times 64$.
Note that, unlike in Section~\ref{sec:f0q2max}, we are now including
finite spatial momentum, so disagreement between the results on the
two different volumes is not an indication of finite volume effects in
this case.

Before we can extract $f_+(q^2)$ from Eq.~(\ref{eq:Fpp}), we need to
calculate $\xi(q^2) = f_-(q^2)/f_+(q^2)$.
This is achieved by constructing a third ratio:
\be
R^3_k(t',t;\vec{p}\,',\vec{p}) = \frac
{C^{K\pi}_k(t',t;\vec{p}\,',\vec{p})C^{KK}_4(t',t;\vec{p}\,',\vec{p})}
{C^{K\pi}_4(t',t;\vec{p}\,',\vec{p})C^{KK}_k(t',t;\vec{p}\,',\vec{p})}
\quad (k=1,2,3)\ .
\label{eq:ratio3}
\ee
We then obtain $\xi(q^2)$ from
\be
\xi(q^2) = \frac
{-(E_K(\vec{p}) + E_K(\vec{p}\,'))  (p+p')_k + 
  (E_K(\vec{p}) + E_\pi(\vec{p}\,'))(p+p')_k R^3_k}
{ (E_K(\vec{p}) + E_K(\vec{p}\,'))  (p-p')_k - 
  (E_K(\vec{p}) - E_\pi(\vec{p}\,'))(p+p')_k R^3_k}\ .
\label{eq:xi}
\ee

We observe that $R^3_k(t',t;\vec{p}\,',\vec{p})=1$ in the
$SU(3)_\text{flavour}$ symmetric limit, and 
deviates only slightly from unity at our simulation quark masses.
Consequently, from Eq.~(\ref{eq:xi}), $\xi(q^2)$ has a small magnitude
$\ltorder 0.1$ with an error typically 25\%-100\%.

Finally, we can double the number of available $q^2$ values by
repeating the steps above (Eq.~(\ref{eq:ratio2})-(\ref{eq:xi})) for
the $\pi\to K$ matrix element as described in Section~V of
Ref.~\cite{Dawson:2006qc}.

\vspace*{-2mm}
\section{Results}
\vspace*{-1mm}
\subsection{Interpolation to $q^2=0$}

\begin{figure}[tb]
\vspace*{-1mm}
\includegraphics[width=7.5cm]{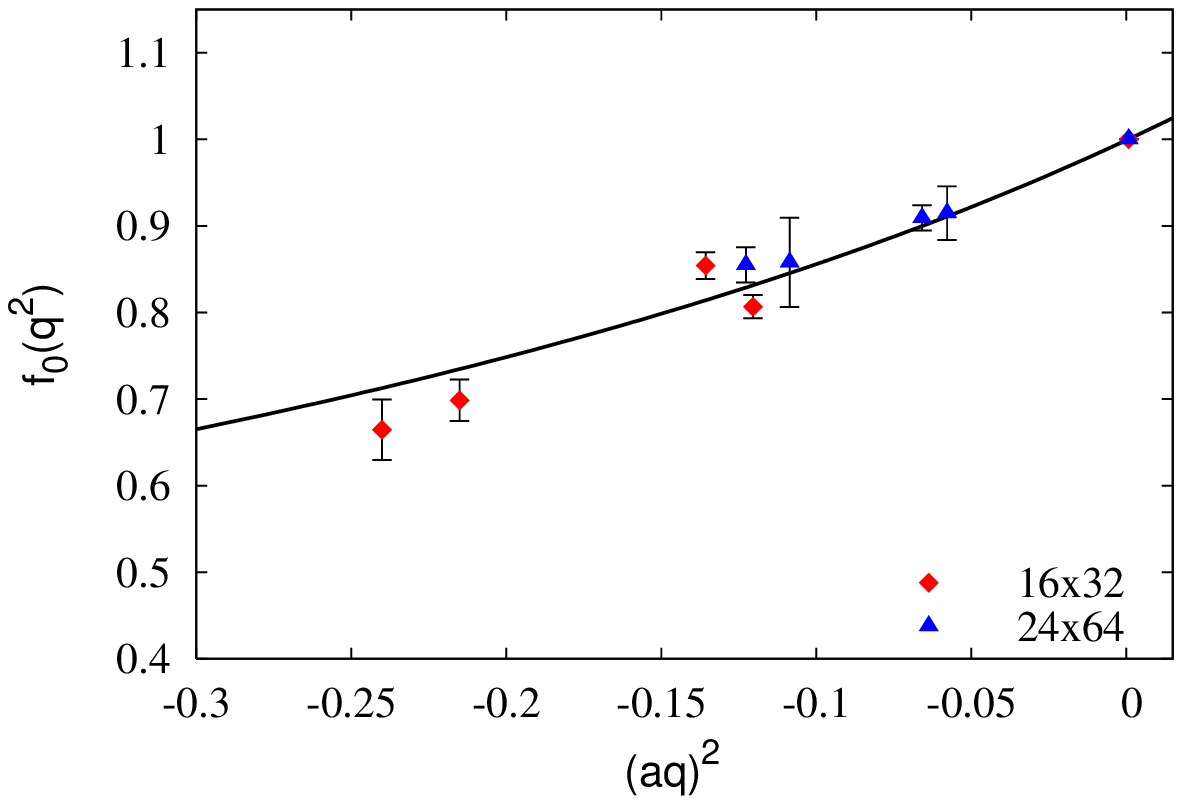}
\includegraphics[width=7.5cm]{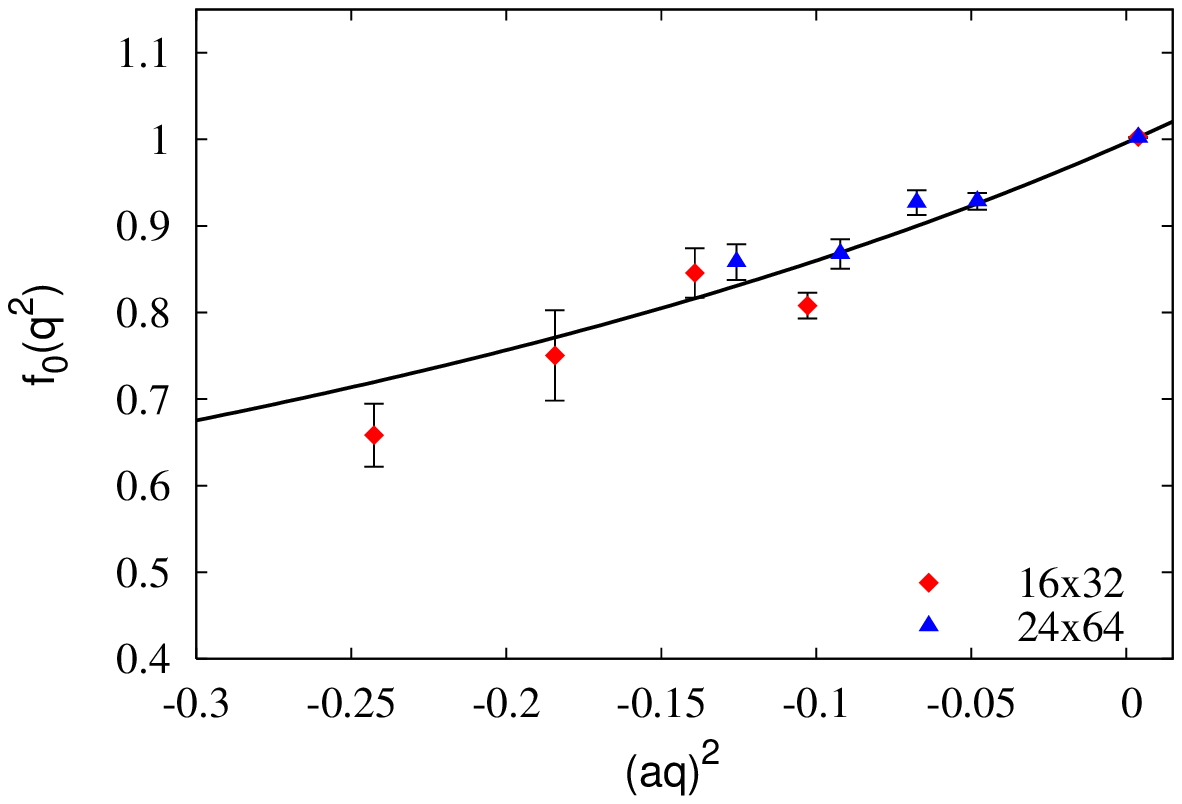}
\vspace*{-8mm}
\caption{Scalar form factor $f_0(q^2)$ for bare quark masses
  $am_{ud}=0.03$ (left) and 0.02 (right). Results are obtained on two
  volumes $V=16^3\times 32$ (red diamonds) and $V=24^3\times 64$ (blue
  triangles). The solid line in each plot is the result of a fit using
  a monopole ansatz (Eq.~({\protect\ref{eq:monopole}})).}
\label{fig:fzero}
\vspace*{-2mm}
\end{figure}

We are now in a position to combine the results obtained above for the
$f_0(q^2_\text{max}),\, F(p,p')$ and $\xi(q^2)$ to reconstruct the
scalar form factor 
\be
f_0(q^2)=f_+(q^2)\left[ 1+ \frac{q^2}{m_K^2 - m_\pi^2}\xi(q^2) \right]
\ .
\ee
We present our results obtained on each volumes for $f_0(q^2)$ in
Fig.~\ref{fig:fzero} for quark masses $am_{ud}=0.03$ (left) and
$am_{ud}=0.02$ (right).
In the intermediate $q^2$-range in both plots, we see good agreement
between the results obtained on the two different volumes, indicating
that finite size effects are negligible, at least for the quark masses
considered here.
This means that we now have results over a large range of $q^2$ to fit
to.

We fit our data with a monopole ansatz
\be
f_0(q^2) = \frac{f_0(0)}{(1-q^2/M^2)}\ ,
\label{eq:monopole}
\ee
which we find to describe our data very well.
This enables us to interpolate our results for $f_0(q^2)$ and
$f_0(q^2_\text{max})$ to $q^2=0$.
Future work will involve testing several ans\"atze in order to obtain
$f_0(0)$, although previous work suggests that results at the
quark masses considered here are insensitive to the choice of
interpolating method.

\vspace*{-1mm}
\subsection{Chiral Extrapolation}
\vspace*{-1mm}

Now that we have obtained results for $f_+(0)=f_0(0)$ at three
different quark masses, we are in a position to attempt an
extrapolation to the physical pion mass.
Inserting these results into the expression given in
Eq.~(\ref{eq:deltaf}), together with $f_2$ calculated at the simulated
quark masses using the ChPT formula
\cite{Leutwyler:1984je,Becirevic:2005py}, we are now left with the 
task of chirally extrapolating $\Delta f$.

The Ademollo-Gatto Theorem implies that $\Delta f \propto (m_s -
m_{ud})^2$, hence we attempt a chiral extrapolation using
\be
\Delta f = a + B(m_s - m_{ud})^2\ .
\label{eq:chiral1}
\ee
Note that in the $SU(3)_\text{flavour}$ limit, $\Delta f=0$, so we
expect that a fit to our data should produce $a\approx 0$.

\begin{figure}[tb]
\vspace*{-1mm}
\includegraphics[width=7.5cm]{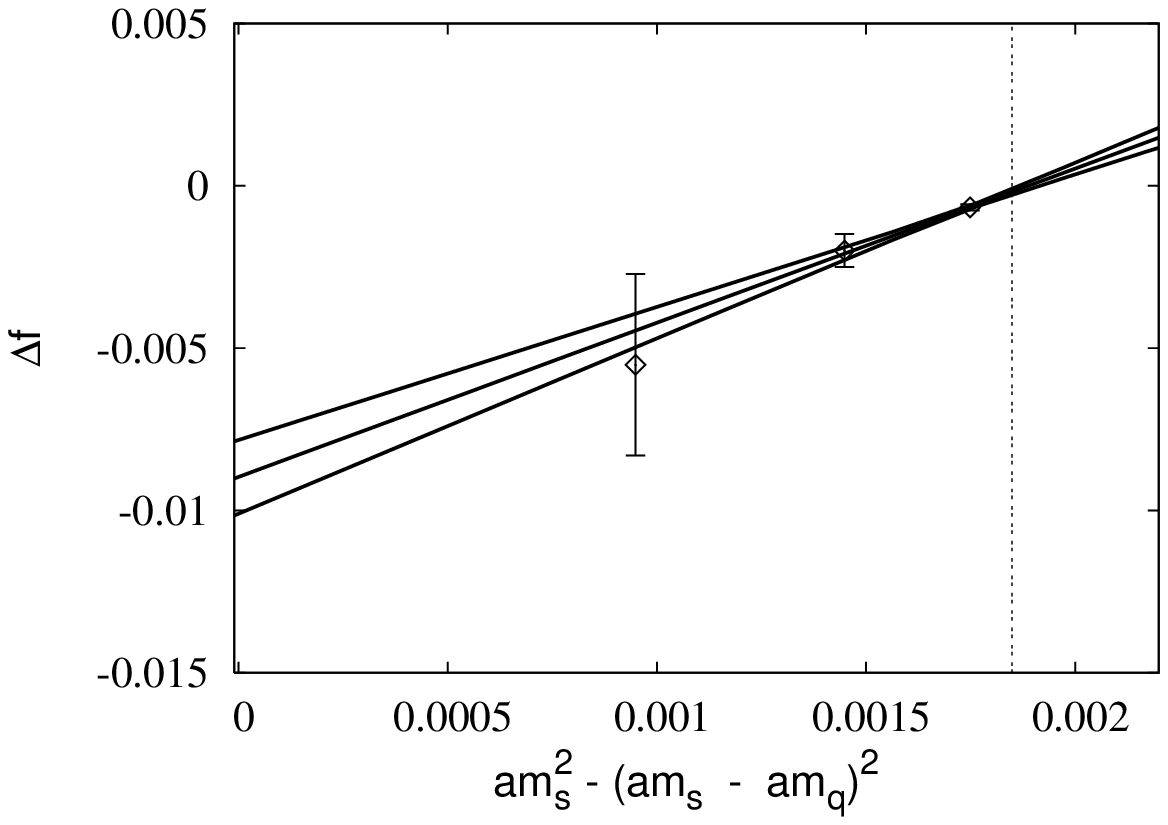}
\includegraphics[width=7.5cm]{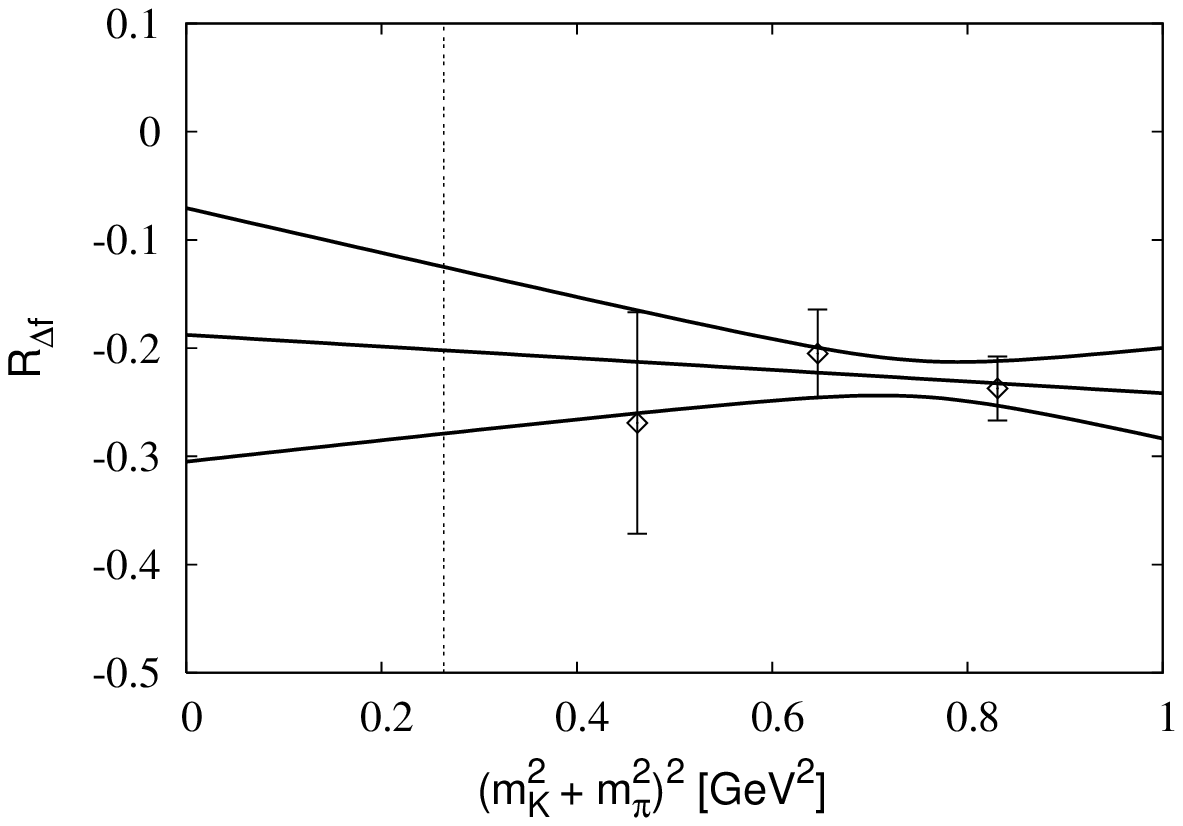}
\vspace*{-8mm}
\caption{The left plot shows the chiral extrapolation of $\Delta f$
  using a trivial modification of Eq.~({\protect\ref{eq:chiral1}}).
  The vertical dotted line indicates the $SU(3)_\text{flavour}$ limit.
  The right plot is an alternative chiral extrapolation
  (Eq.~({\protect\ref{eq:chiral2}}))
}
\label{fig:chiral}
\vspace*{-2mm}
\end{figure}

In the left plot of Fig.~\ref{fig:chiral}, we show the chiral
extrapolation of our results using a slightly modified version of
Eq.~(\ref{eq:chiral1}) which allows the result in the chiral limit to
be obtained from the intercept with the y-axis.
We are encouraged by the fact that $\Delta f$ passes through zero at
the $SU(3)_\text{flavour}$ symmetric point (denoted by the vertical
dotted line), and we find in the chiral limit $\Delta f =
-0.0090(11)$.

Alternatively, it has been noted that is convenient to consider an
extrapolation of the ratio \cite{Becirevic:2004ya,Dawson:2006qc}
\be
R_{\Delta f} = \frac{\Delta f}{(m_K^2 - m_\pi^2)^2} = a + b(m_K^2 +
m_\pi^2)\ .
\label{eq:chiral2}
\ee
Extrapolating our data using this form provides an estimate of the
systematic error in the choice of chiral extrapolation
(\ref{eq:chiral1}).
This obviously requires further investigation and will be reported on
in a forthcoming publication.
We show the extrapolation using Eq.~(\ref{eq:chiral2}) in the right
plot of Fig~\ref{fig:chiral} from which we extract a result at the
physical meson masses (vertical dotted line). 
We take the difference between the results obtained from the two
extrapolations (0.0011) as an estimate of the systematic error due to
the chiral extrapolation.

Finally, since we only have results at one lattice spacing, we are
unable to extrapolate to the continuum limit.
However, lattice artefacts are formally of ${\cal
  O}(a^2\Lambda^2_\text{QCD})\approx 2.5\%$.
Hence our preliminary result is
\be
\Delta f = -0.0090(11)(11)(2) \Rightarrow f_+^{K\pi}(0)=0.9680(16)\ ,
\ee
where the first error is statistical, and the second and third are
estimates of the systematic errors due to the chiral extrapolation and
lattice aritefacts, respectively.
This result agrees very well with a recent two-flavour result
\cite{Dawson:2006qc}, also obtained with domain wall fermions at a
similar lattice spacing $(a^{-1}\approx 1.64\,\text{GeV})$, indicating
that the effects due to a dynamical strange quark are small.

Using $|V_{us}f_+(0)|=0.2169(9)$ from the experimental decay amplitude
\cite{blucher06}:
\be
|V_{us}| = 0.2241(9)_\text{exp}(4)_{f_+(0)}
\ee
we find
$
|V_{ud}|^2 + |V_{us}|^2 + |V_{ub}|^2 = 1-\delta,\quad \delta=0.0015(7)
$
which can be compared with result given by PDG(2006)
$\delta=0.0008(11)$.

\vspace*{-2mm}
\section{Summary and future work}
\vspace*{-3mm}

We have presented a preliminary result for $\Delta f= f_+(0) - (1 +
f_2)$ using $N_f=2+1$ dynamical domain wall fermions with three
choices for the light quark masses.
Our result $\Delta f = -0.0090(11)(11)(2)$ agrees very well with the
$N_f=2$ result \cite{Dawson:2006qc} and confirms the trend of other
lattice results
\cite{Becirevic:2004ya,Okamoto:2004df,Tsutsui:2005cj,Dawson:2005zv}
which prefer a negative value for $\Delta f$, in agreement with the
early result of Leutwyler \& Roos \cite{Leutwyler:1984je}.
We performed our simulations with matched parameters on two volumes
and we observe no obvious finite size effects.

This result can be improved by decreasing the error on the point at
$am_{ud}=0.01$ and simulating at lighter quark masses.
Additionally, this result has been obtained at a single value of the
lattice spacing, so future simulations will need to be performed at
least at one more lattice spacing to investigate scaling behaviour.

\vspace*{-2mm}
\section*{Acknowledgements}
\vspace*{-3mm}
This work is supported under PPARC grants PP/C504386/1 and
PP/D000238/1.
%
%
We thank Peter Boyle, Dong Chen, Mike Clark, Norman Christ,
Saul Cohen, Calin Cristian, Zhihua Dong, Alan Gara, Andrew Jackson,
Balint Joo, Chulwoo Jung, Richard Kenway, Changhoan Kim,
Ludmila Levkova, Xiaodong Liao, Guofeng Liu, Robert Mawhinney,
Shigemi Ohta, Konstantin Petrov, Tilo Wettig and Azusa Yamaguchi
for developing the QCDOC machine and its software.  
This development and the resulting computer equipment used in this
calculation were funded by the U.S. DOE grant DE-FG02-92ER40699, PPARC
JIF grant PPA/J/S/1998/00756 and by RIKEN.
Wish to thank the staff in the Advanced Computing Facility in the
University of Edinburgh for their help and support for this research
programme.


\vspace*{-2mm}
\begin{thebibliography}{99}
\vspace*{-2mm}

\bibitem{Cabibbo:1963yz}
  N.~Cabibbo,
  Phys.\ Rev.\ Lett.\  {\bf 10}, 531 (1963);
%
  M.~Kobayashi and T.~Maskawa,
  Prog.\ Theor.\ Phys.\  {\bf 49}, 652 (1973).

\bibitem{Ademollo:1964sr}
  M.~Ademollo and R.~Gatto,
  Phys.\ Rev.\ Lett.\  {\bf 13}, 264 (1964).

\bibitem{Leutwyler:1984je}
  H.~Leutwyler and M.~Roos,
  Z.\ Phys.\ C {\bf 25}, 91 (1984).

\bibitem{Cirigliano:2005xn}
  V.~Cirigliano {\it et al.},
  JHEP {\bf 0504}, 006 (2005)
  [arXiv:hep-ph/0503108].

\bibitem{Becirevic:2004ya}
  D.~Becirevic {\it et al.},
  Nucl.\ Phys.\ B {\bf 705}, 339 (2005)
  [arXiv:hep-ph/0403217];
%
  D.~Becirevic {\it et al.},
  Eur.\ Phys.\ J.\ A {\bf 24S1}, 69 (2005)
  [arXiv:hep-lat/0411016].

\bibitem{Okamoto:2004df}
  M.~Okamoto  [Fermilab Lattice Collaboration],
  arXiv:hep-lat/0412044.

\bibitem{Tsutsui:2005cj}
  N.~Tsutsui {\it et al.}  [JLQCD Collaboration],
  PoS {\bf LAT2005}, 357 (2006)
  [arXiv:hep-lat/0510068].

\bibitem{Dawson:2005zv}
  C.~Dawson {\it et al.}, 
  PoS {\bf LAT2005}, 337 (2006)
  [arXiv:hep-lat/0510018].

\bibitem{Dawson:2006qc}
  C.~Dawson {\it et al.}, 
  arXiv:hep-ph/0607162.

\bibitem{Iwasaki:1985we}
  Y.~Iwasaki,
  Nucl.\ Phys.\ B {\bf 258}, 141 (1985);
  Y.~Iwasaki and T.~Yoshie,
  Phys.\ Lett.\ B {\bf 143}, 449 (1984).

  \bibitem{Rob} 
R. Tweedie {\it et al.}, these proceedings, PoS {\bf LAT2006} 096
(2006).

\bibitem{Kaplan:1992bt}
  D.~B.~Kaplan,
  Phys.\ Lett.\ B {\bf 288}, 342 (1992)
  [arXiv:hep-lat/9206013];
  Y.~Shamir,
  Nucl.\ Phys.\ B {\bf 406}, 90 (1993)
  [arXiv:hep-lat/9303005].

\bibitem{Hashimoto:1999yp}
  S.~Hashimoto {\it et al.}, 
  Phys.\ Rev.\ D {\bf 61}, 014502 (2000)
  [arXiv:hep-ph/9906376].

\bibitem{Becirevic:2005py}
  D.~Becirevic, G.~Martinelli and G.~Villadoro,
  Phys.\ Lett.\ B {\bf 633}, 84 (2006)
  [arXiv:hep-lat/0508013].

\bibitem{blucher06}
  E.~Blucher and W.J.~Marciano,
  ``$V_{ud},\,V_{us}$, the Cabibbo angle and CKM unitarity'', PDG,
  2006.

\end{thebibliography}
\end{document}